\documentclass[11pt,twoside]{article}
\usepackage{asp2004}
\usepackage{epsfig} 
\usepackage{psfig}
\usepackage{lscape}
\begin{document}
\title{Observations of Active Galactic Nuclei with Ground-Based Cherenkov Telescopes}%%% Fill in title
\author{Henric Krawczynski}   %%% Fill in author names
\affil{Washington University in St. Louis, Physics Department, 1 Brookings 
Drive, CB 1105, St. Louis, MO 63130}    %%% Fill in author affiliations

\begin{abstract} %%% Abstract to run on from here.
Imaging Atmospheric Cherenkov Telescopes (IACTs) allow us to observe
Active Galactic Nuclei (AGNs) in the 100 GeV to 20 TeV energy range with
high sensitivity. The TeV $\gamma$-ray observations of the 
ten blazars detected so far in this energy range reveal rapid flux 
and spectral variability on time scales of several hours, sometimes 
even on time scales of a few minutes. While simple synchrotron-Compton 
models can explain the observed non-thermal emission, alternative 
models which involve high-energy protons are not yet ruled out.
After reviewing the status of the major IACT experiments,
we describe some recent observational results and their 
astrophysical implications. We conclude with a discussion 
of possible avenues for future research.
\end{abstract}
%%% MAIN BODY OF TEXT GOES HERE. CONSULT "INSTRUCTIONS FOR AUTHORS USING
%%% LATEX2E MARKUP", SECTIONS 2.3-2.6 FOR HELP WITH EQUATIONS, FIGURES,
%%% AND TABLES.

%\section{}   %%% Top level section head (remove "%" symbol)
%\subsection{}   %%% Second level section head (remove "%" symbol)
%\subsubsection{}   %%% Lowest level section head (remove "%" symbol)
%\section*{}	%%% Unnumbered top level section head (remove "%" symbol)
%\subsection*{}   %%% Unnumbered second level section head (remove "%" symbol)
\vspace*{-2.5cm}
\section{Introduction}
The EGRET {\it (Energetic Gamma Ray Experiment Telescope)} detector on 
board of the {\it Compton Gamma-Ray Observatory} discovered strong 
MeV $\gamma$-ray emission from 66 Active Galactic Nuclei (AGNs), 
mainly from Flat Spectrum Radio Quasars and Flat Spectrum Radio Sources  \cite{Hart:99}.
As of  this writing (August 2005), ground-based Cherenkov 
telescopes discovered TeV $\gamma$-ray emission from eleven Active Galactic Nuclei, 
only two of which were listed in the third EGRET source catalog (see Table \ref{detect} below). 
Ten of the eleven sources are blazars (nine BL Lac objects and the quasar H 2356-309)
and combine a relatively low luminosity with Spectral Energy Distributions (SEDs) 
that peak at extremely high energies. The eleventh source is associated with the 
FR~I radio galaxy M~87. The GeV/TeV emission from M~87 may originate in
a qualitatively different way than in the blazar type objects, 
and we limit the discussion to the latter source class.

The TeV $\gamma$-ray emission from blazars is believed to 
originate from highly relativistic plasma outflows (jets) emanating
from mass accreting black holes with the jets pointing at us.
The rapid large amplitude flux variability on time scales of 
several minutes \cite{Gaid:96} 
suggests that the emission originates from small regions 
with diameters on the order of 10$^{15}$~cm, less 
than one parsec away from the central engine.
The $\gamma$-ray observations allow us to probe the structure of AGN 
jets very close to the central engines and thus to 
gain key insights into the processes of accretion onto 
a supermassive black hole and jet formation. 
While we focus here on blazars with TeV emission (see also \cite{Kraw:02rev,
Tave:04}), 
reviews on observations and models of sources with  
MeV/GeV emission can be found 
in \cite{Siko:01b,Copp:99}. Broader overviews of the 
field of TeV $\gamma$-ray astronomy are given in
\cite{Buck:02,Ong:03,Trev:03,Feli:04}.
\section{Status of the Major Imaging Atmospheric Cherenkov Telescope Experiments}
\label{exp}
\begin{figure}[t]
\begin{minipage}{6.8cm}
\epsfig{file=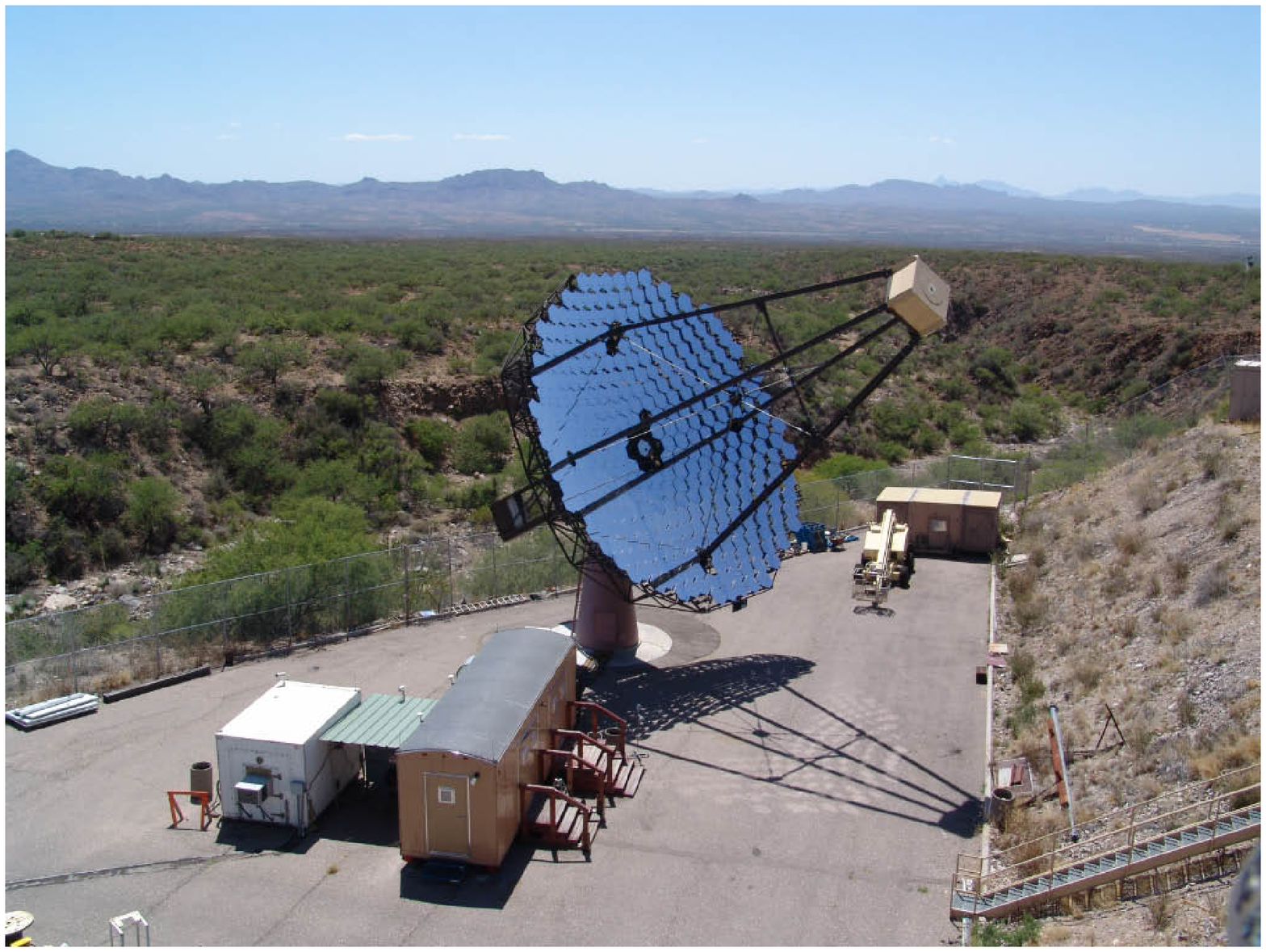,height=4.9cm}
\end{minipage}
\begin{minipage}{6.8cm}
\epsfig{file=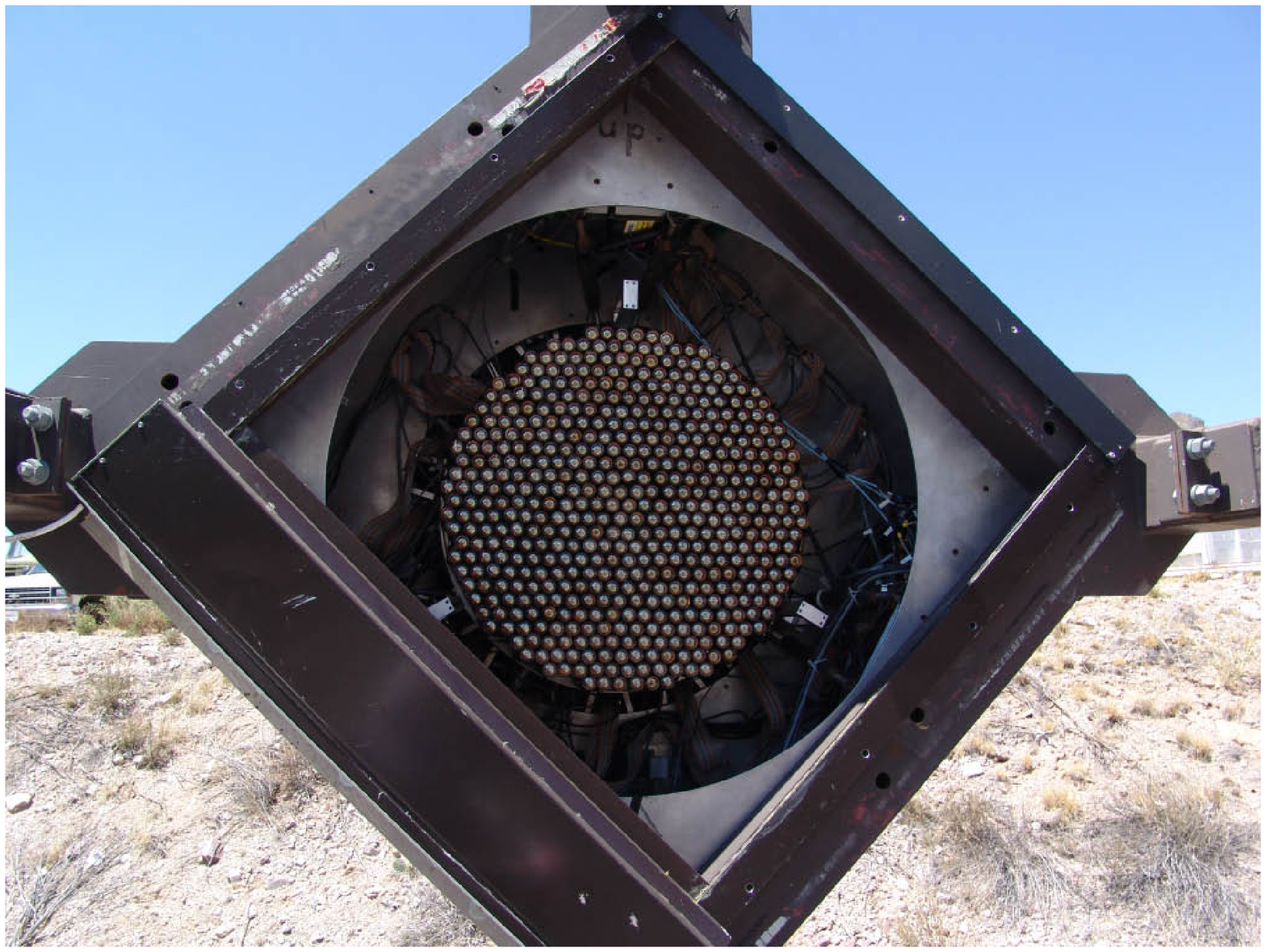,height=4.9cm}
\end{minipage}
\caption{\label{photos} The first of four VERITAS telescopes (left image) and its 499 pixel photomultiplier 
camera (right image) \cite{Hold:05}.}\vspace*{-0.7cm}
\end{figure}
The technique of detecting $\gamma$-rays with Imaging Atmospheric Cherenkov Telescopes (IACTs) 
was pioneered by the Whipple collaboration with the 10 m Whipple telescope. Using a 
fast pixilated camera, the collaboration succeeded in discovering the first galactic 
TeV $\gamma$-ray source, the Crab Nebula between 1986 and 1988 \cite{Week:89}, 
and the first extragalactic TeV $\gamma$-ray source, the BL Lac object Mrk 421, in 1992 \cite{Punc:92}.
While the Whipple collaboration pioneered the use of an imaging camera, the
HEGRA (High Energy Gamma-Ray Astronomy) collaboration showed that an array of 
telescopes operated as a single detector can suppress the dominant Cosmic Ray and muon 
induced backgrounds very effectively and can achieve a very high sensitivity \cite{Daum:97}.

The successor experiment of the Whipple 10~m is the Very Energetic 
Radiation Imaging Telescope Array System (VERITAS) consisting of 
four (eventually seven) 12 m diameter Cherenkov telescopes \cite{Week:02}.
The first VERITAS telescope has been taking data since April 
2004 (Fig.\ 1) and the second will be operational in October 2005 \cite{Hold:05}.
While the first two telescopes have temporarily been assembled at 
Mount Hopkins (AZ), the third and fourth telescopes will be built 
at Kitt Peak (AZ), and the first two telescopes will be moved 
to the latter site to form the full four-telescope array in mid 2006.

The High Energy Stereoscopic Array (H.E.S.S.) \cite{Hint:04} and 
Major Atmospheric Gamma Imaging Cherenkov detector (MAGIC) \cite{Bast:04} 
succeeded the HEGRA and CAT (Cherenkov Array at Themis) \cite{Barr:98} 
experiments. H.E.S.S.\ is an array of four 12~m diameter Cherenkov 
telescopes located in the Khomas Highland in Namibia.
The entire telescope array has been fully operational since December 2003 
and has already produced spectacular science results, e.g., 
the detection of a sample of galactic sources in the direction 
of the Milky Way disk \cite{Ahar:sc}. MAGIC presently consists 
of a single telescope of 17~m diameter located on the Canary Island 
La Palma (Spain). A second 17~m telescope is under construction.
CANGAROO III consists of four telescopes of the 
10~m class located in Woomera, Australia \cite{Kawa:01}. 

The H.E.S.S.\ experiment has demonstrated that the new generation of
Cherenkov telescopes can achieve sensitivities on the order of
0.5\% Crab for 50 hr integration times together with 
single photon angular and energy resolutions of 
0.1$^\circ$ and 15\%, respectively.
\section{TeV Gamma-Ray Observations of Blazars}
\label{obs}
\begin{table}[t]
\begin{center}
\begin{tabular}{|c|c|c|c|} \hline
Source & $z$ & Reference & EGRET Detection \\
 &  & &  \cite{Hart:99} \\ \hline
 M~87        & 0.004  & Aharonian et al.\ 2003a  & no \\
 Mrk 421     & 0.031  & Punch et al.\ 1992 & yes \\
 Mrk 501     & 0.034  & Quinn et al.\ 1996    & no \\
 1ES 2344+514    & 0.044  & Catanese et al.\ 1998 & no \\
 1ES 1959+650    & 0.047  & Nishiyama et al.\ 1999 & no \\
 PKS 2005-489    & 0.071  & Aharonian et al.\ 2005a& no \\ 
 PKS 2155-304    & 0.116  & Chadwick et al.\ 1999  & yes \\
 H 1426+428      & 0.129  & Horan et al.\ 2002     & no  \\
 H 2356-309      & 0.165  & Pita et al.\ 2005  & no \\ 
 1ES 1218+304    & 0.182  & Meyer et al.\ 2005 & no \\
 1ES 1101-232    & 0.186  & Tluczykont et al.\ 2005 & no \\ \hline
\end{tabular}
\vspace*{-2ex}
\end{center}
\caption{Highly significant detections of Active Galactic Nuclei 
by ground based GeV/TeV $\gamma$-ray telescopes (as of August 2005).}
\label{detect}
\vspace*{-2ex}
\end{table}
Since the discovery of TeV $\gamma$-ray emission from the BL Lac object 
Mrk 421 in 1992, TeV emission from other blazars has been 
searched for intensively. The result of this search has been the 
detection of TeV emission from the ten blazars listed 
in Table \ref{detect}. The redshifts of the detected sources 
range from 0.031 for Mrk 421 to 0.186 for 1ES~1101-232.

The TeV $\gamma$-ray fluxes are highly variable and are frequently too low to warrant the 
detection on time scales of a few days. 
The strongest flares have been observed from the sources Mrk 421
and Mrk 501 with flux levels of about 10 times the flux from the 
Crab Nebula (e.g.\ Gaidos et al.\ 1996).
The three sources Mrk 421, Mrk 501, and 1ES 1959+650 have shown strong flaring 
activity for epochs between several weeks and several months.
Interpreting the flaring activity as a noise process, 
these long flaring phases show that this noise process has considerable 
power at low frequencies, and is thus similar to the ``red'' noise 
processes in X-ray binaries \cite{Uttl:05}.

Energy spectra have been measured for all blazars except for  1ES 1218+304. 
Fitting simple power law models $dN/dE\propto E^{-\Gamma}$, photon indices between 
$\Gamma\,=$ 2 and $\Gamma\,=$ 4 have been reported. 
Significant variability of the TeV $\gamma$-ray energy spectrum 
has been detected for the sources Mrk 421 \cite{Kren:02,Ahar:02b} 
and Mrk 501 \cite{Djan:99}.
For Mrk 421, spectral variability with $\Delta\Gamma\sim$1
has been found within a few hours \cite{Kren:var2,Ahar:02b}. 
The TeV flux and spectral hardness seem to be correlated in the sense that 
higher fluxes are accompanied by harder energy 
spectra. The energy spectra of Mrk 421 and Mrk 501 
can be described by power law models with exponential 
cut-offs $dN/dE\propto E^{-\Gamma}\exp{(-E/E_0)}$.
While cut-off energies $E_0$ between 2.5 and 6 TeV have been reported,
power law models with different $\Gamma$-values but with the 
same value of $E_0$ can fit the data from both sources (see \cite{Kren:var} and references therein).
Combining the data from several experiments, the H~1426+428 spectrum 
exhibits a pronounced kink at $\sim$1 TeV that is commonly attributed to
the effect of extragalactic absorption \cite{Ahar:1426}.

The acquisition of good multiwavelength data sets has encountered substantial 
difficulties as the TeV observatories require flares for sampling the 
TeV light curves on a time scale of hours. 
Some sources were observed with excellent multiwavelength 
coverage but during relatively unspectacular quiescent phases; 
in other cases, the sources were flaring, but the fluxes were only 
poorly sampled in frequency space and in time.
The most remarkable result from the multiwavelength campaigns is that 
there is good evidence for a correlation between the X-ray fluxes and 
the TeV $\gamma$-ray fluxes for the two sources 
Mrk 421 \cite{Buck:96,Taka:96,Taka:00,Blaz:05} (see also Fig. 2) 
and Mrk 501 \cite{Djan:99,Samb:00,Kraw:02}. 
The correlation shows considerable scatter (see Fig.\ \ref{corr}) and
for the source 1ES 1959+650 evidence for a TeV flare 
without an X-ray counterpart has been found  \cite{Kraw:04}.

\begin{figure}[t]
%\hspace*{-1cm}
\begin{minipage}{7.1cm}
\epsfig{file=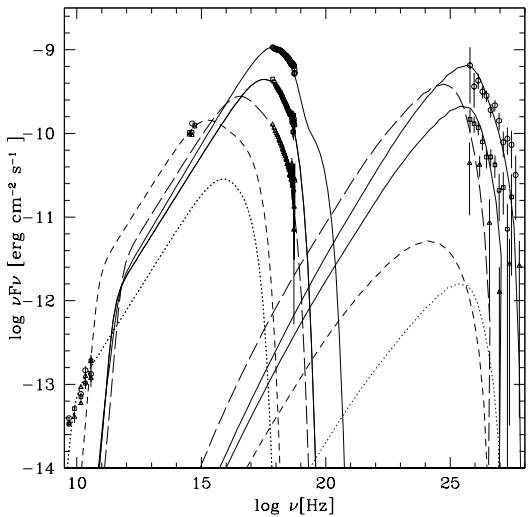,width=7.3cm}
\end{minipage}
\begin{minipage}{7cm}
\caption{\label{sed} Spectral energy distributions of the blazar Mrk 421 at different flux levels (simultaneous
X-ray and TeV $\gamma$-ray data). Synchrotron Self-Compton models for the emission from different zones are 
also shown. The 
% radio data are from the Very Large Array, the optical from the Fred Lawrence Whipple Observatory
% and the Boltwood observatory, the 
X-ray data are from the {\it RXTE} satellite, and the TeV data from the Whipple 10 m telescope.
See Blazejowski et al.\ (2005) for a detailed description of the data and models.
}
\end{minipage}
\end{figure}
The spectral evolution of sources during individual flares might be a powerful tool
to access the elemental processes of Fermi particle acceleration, radiative and adiabatic 
cooling, and diffusive escape from the emission region (Kirk \& Mastichiadis 1999).
X-ray and recently also TeV $\gamma$-ray data do show the predicted signatures:
during flares the sources seem to go through clockwise and counter-clockwise loops in the X-ray or 
$\gamma$-ray hardness--intensity planes \cite{Taka:96,Samb:00,Falc:04}.
However, single sources show a wide range of different behaviours \cite{Taka:00},
indicating that the relative length of the characteristic time scales of particle acceleration and 
radiative cooling change from flare to flare. 
The lack of a prevailing signature has cast doubts on whether the signatures of Fermi
particle acceleration and radiative energy losses have really been observed.

No highly significant evidence has yet been found that the radio, infrared, optical 
or UV fluxes are correlated with the X-ray and/or the TeV $\gamma$-ray emission.
See Buckley et al.\ (1996) for suggestive evidence of a correlation 
and Krawczynski et al.\ (2004) and Blazejowski et al.\ (2005)
for observations lacking such evidence.
\begin{figure}[t]
\begin{minipage}{6.5cm}
\epsfig{file=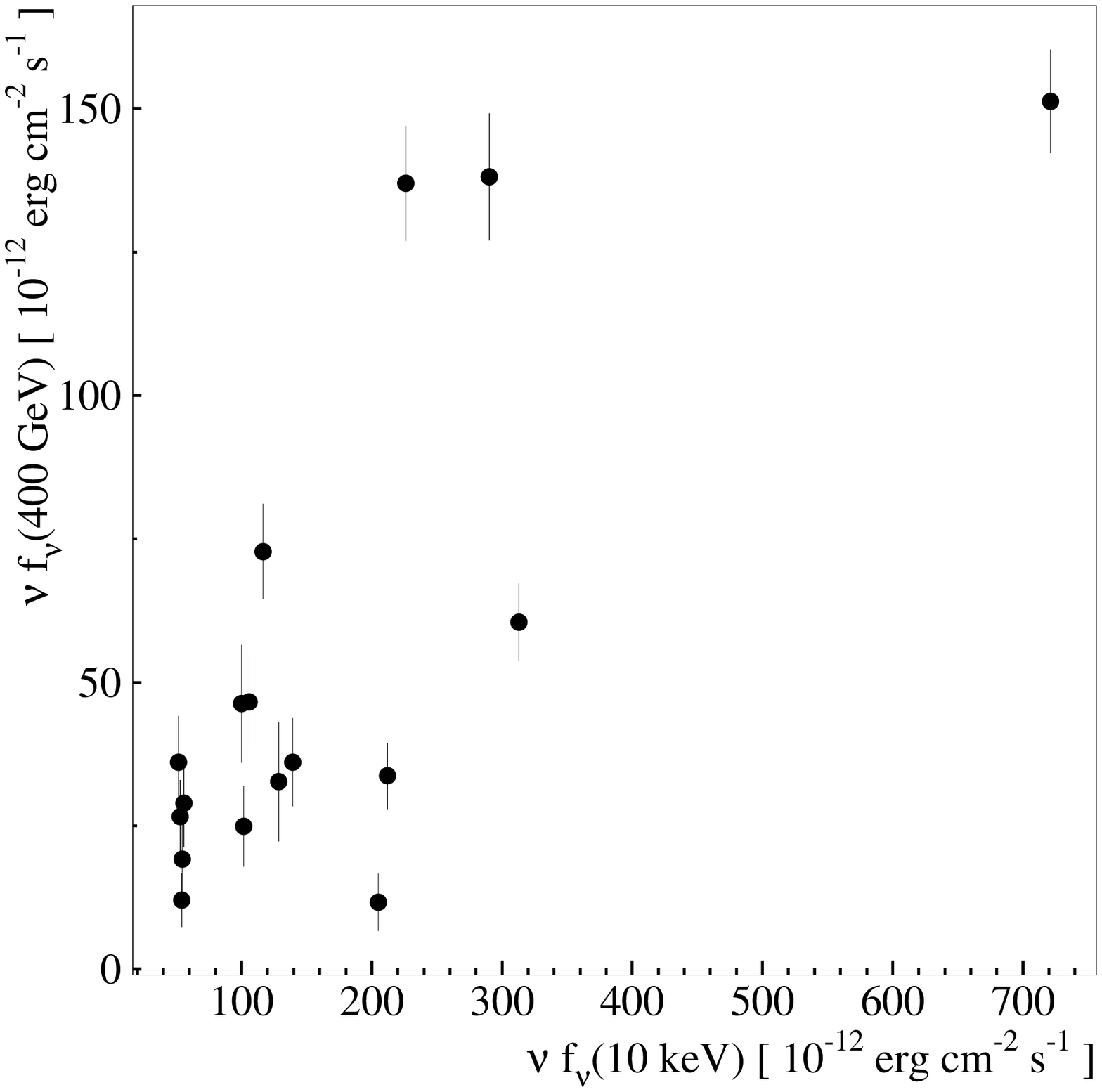,width=6.5cm,height=6.0cm}
\end{minipage}
\begin{minipage}{6.5cm}
\epsfig{file=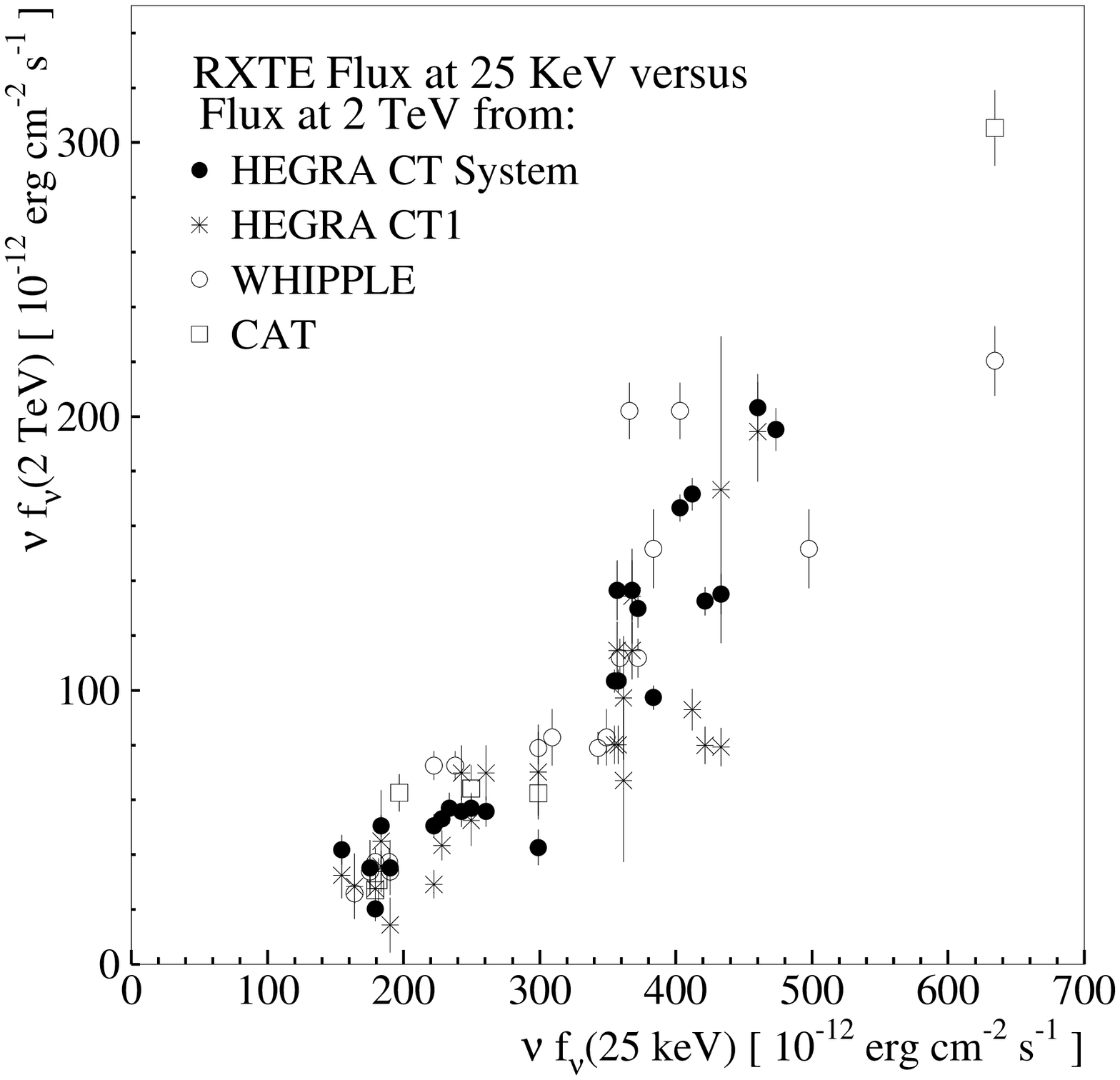,width=6.5cm,height=6.0cm}
\end{minipage}
\caption{\label{corr} Correlation of simultaneous and nearly simultaneous 
X-ray and TeV $\gamma$-ray fluxes from Mrk 421 (left side) and Mrk 501 (right side). 
The data points show fluxes measured in $\sim$20 min long time intervals
spread over $\sim$2 months. The X-ray data are from the {\it RXTE} satellite, 
and the TeV data from the Whipple 10 m telescope (left side) and from CAT, HEGRA, and Whipple
(right side). More details about the observations are given in \cite{Rebi:03,Kraw:02}.}
\end{figure}
\vspace*{-0.35cm}
\section{Recent Development in Interpreting the Data}
\vspace*{-0.35cm}
\label{theory}
The emission from the TeV emitting blazars is commonly attributed to the 
Synchrotron Self-Compton (SSC) process. Embedded in a highly relativistic jet, 
a population of non-thermal electrons (and possibly positrons) emits the radio 
to X-ray emission as synchrotron radiation. The same electron population emits $\gamma$-rays 
through Inverse Compton processes by electrons scattering lower energy 
seed photons. In the case of BL Lac objects, the lack of strong emission lines is 
commonly taken as evidence that ambient photon fields are not important, and that the 
synchrotron photons are the dominant seed photon population.

While simple SSC models can in principle describe most of the experimental data
(see for example Inoue \& Takahara 1996 and Tavecchio et al.\ 2001)
there are some outstanding issues: (i) most models need a minimum Lorentz factor of 
accelerated particles on the order of $\gamma_{\rm min}\,=$ 10$^5$. There is no natural explanation
for the minimum Lorentz factor. (ii) The particle energy density dominates by several 
orders of magnitude over the magnetic field density, making the radiation process 
highly inefficient \cite{Kino:02,Kraw:02}. (iii) While the modeling of the X-ray/TeV $\gamma$-ray 
spectral energy distributions of the TeV blazars requires Bulk Lorentz factors 
$\ge$~25 \cite{Kraw:01a,Kono:03}, direct observations of radio features with the Very Long 
Baseline Array (VLBA) do not show evidence of highly relativistic motion \cite{Pine:04,Pine:05}.

Georganopoulos \& Kazanas (2003) and Ghisellini et al.\ (2005) propose modifications to the
simple SSC model. The first authors postulate that the $\gamma$-ray emission originates
from a fast jet region with electrons that
upscatter low-energy seed photons from the downstream plasma 
that decelerated from bulk Lorentz factors $\sim$15 to $\sim$4. 
The second authors assume that the jet is composed of a fast spine
and a slower envelope.
In this model, electrons in the spine emit the $\gamma$-rays by scattering 
low-energy seed photons from the envelope. 
In both cases, the relative motion of the fast and slow plasma 
components boosts the seed photon densities in the reference frame of the fast components.
As a consequence, the models are able to describe the data with model parameters that
correspond to approximate equipartition between the energy density of relativistic 
electrons (and/or positrons) and the energy density of the magnetic field.

Recently, various authors emphasized that the detailed geometry of the
emission region will impact the observational signatures significantly.
Mimica et al.\ (2005) performed hydrodynamic studies of internal 
shocks in AGN jets arising from collisions of density inhomogeneities. 
Sokolov et al.\ (2005) simulated blazar flares assuming a simple 
geometry of the jet and the particle accelerating shock, 
taking into account light travel delays for the photons acting 
as seed photons for Inverse Compton processes and the 
photons escaping the emission region.
Both studies indicate that simple one-zone models are not able to
adequately describe the detailed observational data that have
become available during the last several years.
It will be a major challenge to find observational signatures 
that constrain the jet physics while they do not depend on 
model details. Examples of such a signature are the soft X-ray 
precursors expected to precede major flares if the jet energy 
is mainly transported by cold electrons and not by 
Poynting flux \cite{Mode:04}.

We have limited the discussion here to leptonic models. 
Discussions of hadronic models can be found in 
(Mannheim et al.\ 1993, Dar \& Laor 1997, Pohl \& Schlickeiser 2000,
Aharonian 2000, M\"ucke et al.\ 2003, Atoyan \& Dermer 2003,
B\"ottcher 2005, Reimer et al.\ 2005).
\section{Outlook}
\label{ol}
The sensitivities of the new experiments H.E.S.S., VERITAS, MAGIC and CANGAROO III surpass or will surpass 
those of the preceding generation (Whipple, HEGRA and CAT) by one order of magnitude.
The highest priority of the upcoming observations is to increase the number of TeV detections 
and to study a larger sample of sources in detail. Using a large statistical sample of sources 
at different redshifts might allow us to constrain the extent to which the GeV/TeV $\gamma$-ray 
energy spectra are modified by extragalactic absorption owing to the GeV/TeV photons 
pair-producing with photons of the Cosmic Infrared Background (CIB). 
The $\gamma$-ray data may give unique information about the intensity and spectrum 
of the CIB and may thus constrain the history of star formation in the early Universe 
\cite{Stec:92,Prim:01,Dwek:04,Dwek:05,Kono:05,Schr:05}. 
The determination of the modification of the $\gamma$-ray energy spectra by extragalactic 
absorption is an important pre-requisite for the astrophysical interpretation of the 
GeV/TeV $\gamma$-ray data. With a better handle on the extent of extragalactic absorption, 
detailed broadband observations should be able to yield an unambiguous identification of 
the emission mechanism.
Once the emission mechanism is known, the matter, energy content, and structure of the 
jets can be constrained, and implications for the accretion and 
jet formation processes can be derived.
\\[0.5ex]
\acknowledgements 
HK would like to thank the VERITAS collaboration for the joint work 
on the VERITAS experiment and the discussions about the astrophysics 
of blazars. He acknowledges gratefully the support by the Department of 
Energy through the Outstanding Junior Investigator award program
and by NASA through the RXTE Guest Observer Program (grant NAG 13770).

\end{document}